\documentstyle[11pt,aaspp4]{article}
\received{}

\lefthead{Demarque et al.}
\righthead{The Metallicity Dependence of RR Lyrae 
Absolute 
Magnitudes from Synthetic Horizontal-Branch Models}

\begin{document}

\title{The Metallicity Dependence of RR Lyrae Absolute 
Magnitudes from Synthetic Horizontal-Branch Models}
\author{Pierre Demarque, Robert Zinn}
\affil{Department of Astronomy, Yale University, P.O. Box 208101, New Haven, 
CT 06520-8101 \\ email: demarque@astro.yale.edu, email: zinn@astro.yale.edu}
 
\author{Young-Wook Lee, Sukyoung Yi\altaffilmark{1}}
\affil{Center for Space Astrophysics, Yonsei University, Seoul 120-749, Korea
 \\ email: ywlee@galaxy.yonsei.ac.kr, yi@srl.caltech.edu}

\altaffiltext{1}{also at California Institute of Technology, 
MC 405-47, Pasadena, CA 91125} 

\begin{abstract}

A grid of synthetic horizontal-branch (SHB) models based on HB
evolutionary tracks with improved physics has been constructed to
reconsider the theoretical calibration of the dependence of
$M_{v}(RR)$ on metallicity in globular clusters, and the slope of the
mean $<\!\!M_{V}(RR)\!\!>$-[Fe/H] relation.
The SHB models confirm
Lee's earlier finding (Lee 1991) that the slope of the
$<\!\!M_{V}(RR)\!\!>$-[Fe/H]
relation is itself a function of
the metallicity range considered (see also Caputo 1997), and 
that in addition, for a given [Fe/H], RR
Lyrae luminosities depend on HB morphology.  This is due to the fact that HB
stars pass through the RR Lyrae instability strip at different evolutionary
stages, depending on their original position on the HB.  
At [Fe/H] = $-1.9$, and
for HB type 0, the models
yield $M_{V}(RR)$ = 0.47 $\pm 0.10$.  The mean slope for the
zero-age HB models is 0.204.  Since there is no simple universal
relation between $M_{v}(RR)$ and metallicity that is applicable to all globular
clusters, the HB morphology of each individual cluster must be taken into
account, in addition to [Fe/H], in deriving the appropriate $M_{v}(RR)$.  
Taking
HB morphology into account, we find that the slope of the mean 
$<\!\!M_{V}(RR)\!\!>$-[Fe/H] relation varies between 0.36 for the clusters
with galactocentric distances $R_{gc}$
less than 6 kpc and 0.22 for clusters with $6<$$R_{gc}$$\leq 20 kpc$.
Implications for interpreting observations of field RR Lyrae variables and for
absolute globular cluster ages and galactic chronology are briefly discussed.

\end{abstract}


\section{Introduction}
	It has long been realized that determination of the 
absolute ages of globular star clusters in our Galaxy is most vulnerable 
to the uncertainty in their distances (see e.g. Renzini 1991).  
Because of the difficulty in determining the precise position of the 
unevolved main sequence in globular clusters, the RR Lyrae variables 
have been in recent years the preferred distance indicators for globular 
clusters, and used in the calibration of the $\Delta$V method for
comparing observed CMD's to theoretical isochrones  
(Renzini 1991; Buonanno et al. 1994).  In this method, the quantity $\Delta$V 
is defined as the difference in V magnitude between the main sequence turnoff 
bluest point and the HB at the same color in the cluster CMD.  This 
approach has the advantage of being independent of distance and  interstellar 
reddening.  The quantity $\Delta$V can in principle be calibrated 
with theoretical isochrones since it relies primarily on the physics of the 
deep interior, which is believed 
to be relatively well-known, and less on uncertain assumptions about the 
efficiency of convection in the envelope and on color transformations 
(see e.g. the discussion of model 
uncertainties by Chaboyer et al. 1996a).

The purpose of this paper is to present a new theoretical calibration of 
the HB luminosity based on updated stellar evolutionary sequences 
and synthetic HB (SHB) models, and to analyze the sensitivity of 
$M_{V}(RR)$ to [Fe/H]. Other parameters, such as HB morphology and input 
parameters used in the construction of SHBs, also affect the relation.
The slope of the $<\!\!M_{V}(RR)\!\!>$-[Fe/H]
relation is needed to discuss the relative ages of globular clusters and 
to determine whether there exists an age-metallicity correlation in the 
Galactic halo (Zinn 1985, 1993; Chaboyer, Demarque \& Sarajedini 1996).  
Both the slope and zero-point (and their uncertainties) are also essential in 
discussions of the absolute ages of the globular clusters 
(Chaboyer et al. 1996, 1998; Demarque 1997).

With the first parallaxes for field subdwarfs from the Hipparcos 
satellite now becoming available, there has been a flurry of new 
interest in using the main sequence to derive globular cluster 
distances independently from the RR Lyrae luminosity calibration 
(Reid 1997; Gratton et al. 1997; 
Salaris, Degl'Innocenti \& Weiss 1997). 
In early studies, the distance modulus of a globular cluster was 
derived by fitting the cluster main sequence to a main sequence 
derived from field subdwarfs with the same metallicity as the cluster 
(e.g. see the review by Sandage 1986).  But because the uncertainties 
in trigonometric parallaxes for field subdwarfs were large, and the 
photometry of globular cluster main sequence stars was very 
uncertain, this approach had generally been abandoned in recent 
years.  Preliminary analyses of the Hipparcos data indicate that the 
subdwarfs are intrinsically more luminous than previously believed, 
and therefore that globular cluster distances have been underestimated 
in the past. If correct, this means a more luminous HB and a higher 
luminosity for the RR Lyrae variables.  In view of the crucial importance 
of globular cluster distances in understanding the 
evolution of galaxies and cosmology, new theoretical SHB models
have been constructed.

\section{Evolutionary Tracks}

The evolutionary sequences presented here are quite similar to the 
many other HB models in the recent literature (Sweigart 1987; Lee \& 
Demarque 1990; 
Dorman, Lee \& VandenBerg 1991; Dorman 1992; 
Yi, Lee \& Demarque 1993; 
Castellani et al. 1994; 
Caputo \& Degl'Innocenti 1995; Mazzitelli, D'Antona \& Caloi 1995), 
except for recent updates in the opacities and 
equation of state (Rodgers 1986; Iglesias \& Rodgers 1996; 
Rodgers, Swenson \& Iglesias 1996).

The evolutionary sequences are adopted from Yi, Demarque \& Kim 
(1997) but extended to a 
finer grids of metallicity and mass, i.e. in the range 0.52 -- $0.92 M_\odot$ 
at $0.02 M_\odot$ intervals, for the following metallicities:
Z = 0.0001, 0.0002, 0.0004, 0.0007, 0.001, 0.002, and 0.004, 
corresponding to [Fe/H] between $-2.3$ and $-0.7$.  The helium content by
mass $Y$ was taken from the evolutionary tracks that were used for the new
Yale Isochrones\footnote{See Yi et al. at http://www.shemesh.gsfc.nasa.gov/
for both HB tracks and isochrones used in this study.}, 
corresponding to an initial $Y = 0.23$. The main difference in
the input physics with the Lee \& Demarque (1990) models is the introduction
of the OPAL opacities and equation of state (Iglesias \& Rodgers 1996;
Rodgers, Swenson \& Iglesias 1996).  
	As a result, for $Y_{MS}=0.23$, the current HB models of 
Yi et al. (1997) used in this study are approximately 0.05 -- 0.1 mag. 
fainter than those of Lee \& Demarque (1990).
	The Yi et al. models are approximately 0.02 -- 0.05 mag. fainter than 
those of Caloi et al. (1997) and Cassisi et al.(1998)
on the zero-age HB (ZAHB) in the instability 
strip at [Fe/H]~$= -2$.
	This difference seems to be caused mostly by the fact that 
the helium core masses of the Yi et al. models are 
smaller, by 2 -- 5\%, than the core masses of the Caloi et al. and 
Cassisi et al. models.
	Table 1 lists the helium abundances in the envelope and the helium 
core masses for given metallicities used in the Yi et al HB models. These 
values are obtained from  
stellar models at the onset of helium ignition at the tip of the giant branch,
which included the effects of the OPAL opacities, and the same equation of
 state as in Guenther et al. (1992).

\section{Synthetic Horizontal-Branch (SHB) Models}

The SHB models were derived using the technique introduced by Rood (1973), 
and extended by Lee et al. (1990).  The mass distribution on the HB is 
defined by  the following truncated Gaussian distribution:
\begin{equation} 
\Psi(M)\, = \,\Psi_{0}\, [\, M\, - \,(\overline{M_{HB}}\,-\,\Delta M)\,]\,
(M_{RG}\,-\,M) \,\, exp\,[\,-\,\frac{(\overline{M_{HB}}\,-\,M)^{2}}{2 \sigma^{2}}]
\end{equation}
where $\sigma$ is a mass dispersion factor in solar mass, 
$\Psi_{0}$ is a normalization factor, and 
$\overline{M_{HB}}$ ($\equiv M_{RG} - \Delta M$) is the mean mass of HB stars. 
	Three values of $\sigma$ (0.01, 0.02 and 0.03 solar masses) have 
been chosen, where the preferred value $\sigma$ = 0.02 is 
the mean of a representative group of clusters (see Table 1 in
Lee 1990), and 0.01 and 0.03 were added to illustrate the sensitivity
of the SHB models to the choice of $\sigma$. 

When considering the properties of stars in the RR Lyrae instability 
strip, it is convenient to introduce the HB Type index, which is 
defined as the ratio (B-R)/(B+V+R), where B, V, and 
R are the numbers of blue HB stars, RR Lyrae variables, and red HB 
stars, respectively.  This parameter is convenient in classifying HB 
morphology; we note that (B-R)/(B+V+R) ranges from $-1$, for clusters 
that display only a red HB (e.g. 47 Tuc), to $+1$, for clusters 
containing only blue HB stars (e.g. NGC 6752).  Clusters that 
contain nearly equal numbers of red and blue HB stars (e.g. M3), are 
assigned values of (B-R)/(B+V+R) near 0.  One of the advantages of this 
morphology index is that it includes the RR Lyrae variables, thus 
distinguishing between two clusters with the same numbers of blue 
or red HB stars, but differing in their RR Lyrae populations.

Figure 1 illustrates the dependence of the mean RR Lyrae visual 
absolute magnitude $<\!\!M_{V}(RR)\!\!>$ on metallicity and on HB
Type, based on the SHB models constructed with $\sigma = 0.02$.  
The colors and bolometric corrections were taken from Green et al. (1987)
As discussed earlier, the use of improved physics reduces the luminosities 
of SHBs for a given helium abundance.
At [Fe/H] = $-1.9$, and
for HB type 0, the models
yield $M_{V}(RR)$ = 0.47 $\pm 0.10$.  This result is consistent with 
Walker's (1992) value for the LMC RR Lyrae
variables which is based on the classical cepheid distance scale, and with
the distance of SN1987a (Panagia et al. 1991; Gould 1995; Sonneborn 1997).

Using observational data, Fusi Pecci et al. (1992) concluded that 
(B-R)/(B+V+R) depends primarily on the location of the peak of the color 
distribution of the HB, and only slightly on the dispersion in color of the 
HB stars.  This conclusion is confirmed by the theoretical SHB 
models.  For a fixed metallicity, the run of $M_{V}(RR)$ 
predicted by the SHB models as a function of HB Type is found 
to depend weakly on the choice of the mass dispersion $\sigma$ in eq. (1).  
This is illustrated in Figure 2 where RR Lyrae magnitudes for two extreme 
values of $\sigma$, 0.01 and 0.03, are plotted.
The peculiar behavior near HB type 0.9 is caused by the particuler 
metallicity dependence of the vertical width of the HB tracks.  This
vertical width is 
narrower for Z = 0.0002 than for Z = 0.0004.  Thus although the 
ZAHB luminosity is brighter for Z = 0.0002, the evolved RR Lyrae 
variables of Z = 0.0004, which are near the end of their HB tracks, could be 
brighter than those for Z = 0.0002.  Increasing $\sigma$ 
will dilute this effect. As a result, we do not see this 
behavior al larger $\sigma$ in Figure~2.   

\section{Is there a Universal Slope to the $<\!\!M_{V}(RR)\!\!>$-[Fe/H] Relation?}

The dependence of $<\!\!M_{V}(RR)\!\!>$ on [Fe/H] is needed to derive the 
relative ages of globular clusters of different metallicities.  It is
critical in deriving the chronology of the Galactic halo, its 
chemical enrichment, and in particular the possible existence of an 
age-metallicity correlation in the halo.  It is customary to assume a 
linear relation between $M_{V}(RR)$ and [Fe/H], i.e. to write:
\begin{equation} 
M_{V}(RR)\, = \mu\, [Fe/H] + \gamma
\end{equation}
where, when used for globular cluster dating, the slope 
$\mu$ affects 
the relative ages of clusters of different metallicities, and both $\mu$ 
and $\gamma$ determine the absolute ages.

There has been much debate about the value of the slope $\mu$ over 
the years.  Recently, from an analysis of the Oosterhoof-Sawyer 
period shift effect in globular clusters, Sandage (1993) has derived a 
``steep'' $\mu = 0.30 \pm 0.12$, while studies based on 
the Baade-Wesselink method of determining the absolute magnitude of variable 
stars have yielded a ``shallow'' slope $\mu$, in the vicinity of 0.20 or 
less; Jones et al. (1992) derived $\mu = 0.16\pm0.03$ and Skillen et 
al.(1993) determined $\mu = 0.21\pm0.05$.  
Recent analyses of these data by Sarajedini et al. (1997) and by 
Fernley et al. (1998) yielded $\mu = 0.22 \pm 0.05$ and $0.20 \pm 0.04$,
respectively.
HST observations of the HB luminosity of three clusters in M31 
(by Ajhar et al. 1996) have  yielded a very shallow value 
of $\mu$ = 0.08 $\pm 0.13$.  
	Also using HST observations of the CMD's of eight globular clusters 
in M31, Fusi Pecci et al. (1996) derived 
$<\!\!M_{V}\!\!>$ = (0.13 $\pm 0.07$)[Fe/H] + (0.95$\pm 0.09$) for the mean
magnitude of the HB in the instability strip.  Theoretical estimates
have consistently yielded $\mu$ values in the range 0.18-0.20 for models
near the ZAHB (Lee et al. 1990; Salaris et al. 1997). 

Discussions of the theoretical $<\!\!M_{V}(RR)\!\!>$-[Fe/H] relation
are frequently made using ZAHB models (e.g., Caloi et al. 1997).
Sometimes an evolutionary correction is applied to take into account
the fact that RR Lyrae variables are not observed in their original
ZAHB position, and have evolved both in color and magnitude (Carney et
al. 1992).  Synthetic HB models are needed to provide a realistic
description of these evolutionary corrections, which are found to
differ significantly depending on the mass and the chemical
composition of the models.  Furthermore, only with SHB models is it
possible to evaluate the effects of HB morphology on the value of
$<\!\!M_{V}(RR)\!\!>$-[Fe/H] for a given metallicity, as was done
originally by Lee (1991).

In his study of the RR Lyrae luminosities in $\omega$ Cen, 
Lee (1991) pointed out that $M_{V}(RR)$ is not a unique function of 
metallicity, particularly at the lowest metallicities.  The effect is 
particularly marked for clusters with very blue stars on the HB, 
corresponding to HB Types approaching +1.  Figure 1 updates the original 
Lee calibration.  It is clear from Figure 1 that there is nothing 
universal about the value of $\mu$, and great caution should be used 
when applying equation (2) to derive RR Lyrae magnitudes without taking into
account the HB morphology type of the population to which they belong 
(see also Caputo 1997).
 
To examine this point in more detail, let us consider first the
hypothetical case where the HB is evenly populated with stars over a
wide range in [Fe/H].  Since there is very small variation in
$<\!\!M_{V}(RR)\!\!>$ over the range in HB 
type -0.5 to +0.5 (see Figs. 1 and 2),
we may use the calculations for HB type = 0.0 to approximate this
case.  The resulting relationship between  $<\!\!M_{V}(RR)\!\!>$
 and [Fe/H] is 
shown in Figure 3, where one can see that there are small variations
in slope over narrow ranges in [Fe/H] (see also Caputo 1997).  The
slopes found from our calculations are listed in Table 2.  Since most
observational studies have considered variables that span a wide range
in [Fe/H], e.g., \,-2.2 to \,-0.5, it is the slope over a wide range
that is of most interest.  While the relationship given by our
calculations is non-linear, the departures from a straight-line fit
are small (see Figure 3) and would be very hard to detect
observationally.  The slope of the line in Figure 3 (0.21) is similar
to that given by ZAHB calculations, but the zero-point (0.89) is somewhat
brighter because the RR Lyrae variables have evolved from the ZAHB.
It is expected that this $<\!\!M_{V}(RR)\!\!>$-[Fe/H] relationship will not
apply to all stellar populations because HB morphology changes with
[Fe/H], the so-called first parameter, and also varies at constant
[Fe/H], the second parameter effect.  This is most easily illustrated
by considering the relationships expected for the HB morphologies of
the globular clusters lying in different radial zones in the Milky
Way.

Previous investigations (e.g., Searle \& Zinn 1978, Lee et al. 1994)
have shown that the globular clusters of the inner halo and bulge
exhibit a tight relationship between HB morphology and [Fe/H], which
may mean that the second parameter effect is absent or weak among
these clusters.  The top graph in Figure 4 shows this relationship for
the globular clusters that have galactocentric distances,
$R_{gc}$$\leq 6 kpc$ (data from the 1999 June 22 revision of the
Harris 1996 catalogue).  
	To derive the $<\!\!M_{V}(RR)\!\!>$-[Fe/H] relationship
for this group of clusters, we estimated the mean HB type of the
clusters at the metallicities of our calculations and then derived
$M_{V}(RR)$ from the synthetic HB for that HB type and [Fe/H].  When
estimating the mean HB type, lower weight was given to the clusters
with HB type $\sim 1.0$, because these very blue HB clusters contain
few, if any, RR Lyrae variables.  To obtain additional points, we
interpolated in [Fe/H] midway between the values of our calculations.
Figure 4 shows the mean HB types (x's) used in this procedure and the
resulting values of $<\!\!M_{V}(RR)\!\!>$ are 
plotted against [Fe/H] in the top
diagram of Figure 5.  Because the globular clusters with [Fe/H] 
$<-1.5$ have exclusively very blue HB types, in this [Fe/H] range
$M_{V}(RR)$ is significantly brighter than the HB type = 0 case (the
dashed line).  As Figure 5 illustrates, this produces a very steep ($\mu
= 0.36$) $<\!\!M_{V}(RR)\!\!>$-[Fe/H] relationship.

There is not a tight relationship between [Fe/H] and HB morphology
among the globular clusters in the outer halo because of the second
parameter effect.  To obtain a sufficiently large sample of cluters to
illustrate this, it is necessary to consider a wide range in $R_{gc}$
because the number density of clusters falls off steeply with
increasing $R_{gc}$.  The clusters having $6<$$R_{gc}$$\leq 20 kpc$
are plotted in the lower diagram of Fig. 4, where the diversity in HB
types among the metal-poor clusters is obviously much larger than in
the inner halo.  Following the same procedures as before, we have
estimated the mean HB types of the clusters at different values of
[Fe/H] and have estimated values of $M_{V}(RR)$ from our synthetic HB
calculations.  The resulting $M_{V}(RR)-[Fe/H]$ relationship is shown
in the lower diagram of Fig. 5.  In contrast with the relationship for
the inner halo, this one ($\mu = 0.22$, $\gamma = 0.90$) deviates only
slightly from the case where HB type = 0 for all [Fe/H] (see Fig.3).

On the basis of Figs. 4 and 5, we conclude that a universal
$<\!\!M_{V}(RR)\!\!>$-[Fe/H] relation does 
not exist because $<\!\!M_{V}(RR)\!\!>$ depends
on HB morphology as well as [Fe/H] and because the relationship
between HB morphology and [Fe/H] varies with the stellar population
being considered.  While this last point is best illustrated by the
globular clusters in the inner and outer halo (Fig. 4), the recent
work on the color-magnitude diagrams of the globular clusters in M31 
(Fusi Pecci et al. 1996),
M33 (Sarajedini et al. 1998), 
the LMC (Olsen et al. 1998), and the Fornax dwarf spheroidal galaxy 
(Buonanno et al. 1998) also show
significant variations in the HB type - [Fe/H] relation from system to
system.  The 
data on these cluster systems, which for
some of them is far from complete, suggest that the inner halo
galactic may be the most extreme example where over a wide range
of [Fe/H] only very blue HB types are found.  Therefore, stellar
populations in which $<\!\!M_{V}(RR)\!\!>$-[Fe/H] relation is as steep as the
one for the inner halo clusters may be rare.  Nonetheless, one should
be cautious when adopting a $<\!\!M_{V}(RR)\!\!>$-[Fe/H] relation without
information on the variation of HB type with [Fe/H] in the stellar
population.

It is important to consider if the debate over the slope and
zero-point of the $<\!\!M_{V}(RR)\!\!>$-[Fe/H] relation 
is partially due to the
non-universality of this relation.  The relation found here for the
inner halo clusters resembles in slope the steep relationships that
Sandage (1993 and references therein) obtained during his more than a
decade long analyses of the Oosterhoff effect among the galactic
globular clusters.  We doubt, however, that they are related.  Most of
the clusters Sandage used in his analysis are rich in RR Lyrae
variables and do not have extremely blue HBs.  Our models predict that
the $<\!\!M_{V}(RR)\!\!>$-[Fe/H] relation for these clusters is not steeply
sloped.  The $<\!\!M_{V}(RR)\!\!>$-[Fe/H] relation 
that Fusi Pecci et al. (1996)
obtained from the color-magnitude diagrams of 8 globular clusters in
M31 represents the other extreme, for they obtained the 
very shallow slope of $\mu = 0.13\pm0.07$.  In only the 3 most metal
poor clusters of this sample is the HB morphology sufficiently blue to
populate the instability strip with RR Lyrae variables, and these
clusters do not have very blue HB types in spite of [Fe/H] $\sim-1.8$
to $\sim-1.5$ (compare with the top diagram of Fig. 4).  The other 5
clusters in this M31 sample have HB morphologies too red for RR Lyrae
variables, and for them Fusi Pecci et al. had to resort to estimating
the HB level at the instability strip from the observed red HB.  Fusi
Pecci et al. point out that their conclusion that $\mu$ is small does
not depend critically on this uncertain procedure.  For this sample of
M31 clusters, our calculations predict a $<\!\!M_{V}(RR)\!\!>$-[Fe/H] 
relation
similar to the HB type = 0 case; hence $\mu\sim0.21$.  While this
value is larger than the value obtained by Fusi Pecci et al. (1996),
it is barely within one standard deviation of it.  As Fusi Pecci et
al. suggest, a larger sample of M31 clusters must be observed
before one can be confident that the slope of the M31 relation is
indeed incompatible with theoretical calculations such as ours.  The
zero-point $\gamma$ of the M31 relation is fixed entirely by the
distance adopted for M31, and the value obtained by Fusi Pecci et
al. (1996) (0.95$\pm0.09$) is consistent with our calculations.

Of course, the $<\!\!M_{V}(RR)\!\!>$-[Fe/H] relation has also been investigated
using samples of field RR Lyrae variables lying within a few
kiloparsecs of the Sun.  Some of these variables are members of the
galactic halo, while others, preferentially the more metal-rich ones,
are members of the thick disk population.  Because it is difficult to
recognize red HB stars in the field, the HB type - [Fe/H] relations of
these populations are poorly known.  The work by Preston, Shectman and
Beers (1991) on the field HB stars and by Caputo (1993) on the RR Lyrae
variables suggests that the HB morphology of the field may vary with
$R_{gc}$ in much the same way as the HB morphology of the globular
clusters.  Since a few metal-poor clusters with HB type $< 0.9$ have
values of $R_{gc}$ that are not much different from the Sun's, we
suspect the HB type - [Fe/H] relation of the field population near the
Sun may resemble more the lower diagram of Fig. 4 than the upper.  If
this is correct, our models suggest that $\mu \sim 0.22$.  This is
close to the results obtained from applying the Baade-Wesselink method
to samples of field RR Lyrae variables (see above).  The values of
$\gamma$ from the Baade-Wesselink analyses, which are considered more
susceptible to systematic error than $\mu$, are about 0.1 mag. fainter
than the value given by our calculations (see Fernley et al. 1998). 

In principle, the application of the method of statistical parallax to
samples of field RR Lyrae variables should also yield the
$<\!\!M_{V}(RR)\!\!>$-[Fe/H] relation.  To date 
the samples of stars have proved
inadequate for determining both $\mu$ and $\gamma$, but precise values of
$<\!\!M_{V}(RR)\!\!>$ have been obtained at the mean [Fe/H] of the sample of
halo RR Lyrae variables.  Recent results by Layden et al. (1996) and
by Gould \& Popowski (1998) have yielded $M_{V}(RR)$ of $+0.71\pm0.12$
and $+0.77\pm0.13$ at [Fe/H]$\sim -1.6$, respectively.  Our
calculations give brighter values with the exact value depending on
which HB type - [Fe/H] relation is adopted.  For the outer halo one,
which may be appropriate for the variables near the Sun, $M_{V}(RR) =
+0.55$ at [Fe/H]$=-1.6$, which slightly more than one sigma brighter
than the results from statistical parallax.  We have no explanation
for this difference, which if confirmed, may mean that some revision
of the models is required.  While the dependence of $M_{V}(RR)$ on HB
morphology may not have a large effect on the either the
Baade-Wesselink or the statistical parallax analyses, it is probably 
a contributing factor to the scatter in $M_{V}(RR)$ among the 
metal-poor variables (see Jones et al. 1992; Fernley et al. 1998).

\section{Discussion}

The dependence of $<\!\!M_{V}(RR)\!\!>$ on HB morphology may have a
significant effect upon astrophysical problems involving the distance scale
for old stellar populations.  We concentrate here on the
question of the ages of globular clusters, which is important for both
galactic evolution and for setting the minimum age of the universe.
Our results suggest that assuming the same
$<\!\!M_{V}(RR)\!\!>$-[Fe/H] relation for clusters of all HB types
will significantly underestimate the distances to the metal-poor
clusters having very blue HB morphologies.  The luminosities of their
main-sequence turnoffs will be underestimated and their
ages will be overestimated.  This effect is greatest for the
clusters having the very bluest HB types ($\sim 1.0$) (see also Storm
et al. 1994; Clement et al. 1999).

To illustrate this, let us consider the globular cluster M92
(NGC 6341), which is one of the most metal poor globular clusters
([Fe/H]=-2.24, Zinn \& West 1984) and perhaps one of the oldest.
While M92 has a blue HB, it is not extremely blue (HB type = 0.88, Lee
et al. 1994), and one might think that it is immune to this systematic
error.  According to our models, for this HB type
$<\!\!M_{V}(RR)\!\!>$ is approximately 0.07 mag. brighter than the HB
type = 0 case.  Therefore, if the distance modulus of M92 is set using
a $<\!\!M_{V}(RR)\!\!>$-[Fe/H] relation that is appropriate for HB
types in the range -0.5 to +0.5, then the luminosity of its
main-sequence turnoff will be underestimated by about 0.07 mag.  This
will cause its age to be overestimated by about 1 Gyr.  This $\leq$
10\% error in cluster age may appear small, it is nonetheless
significant for either setting the lower limit on the age of the
universe or for ascertaining the dispersion in age among the globular
clusters and thereby distinguishing between different scenarios for
the formation of the galactic halo.

The theoretical prediction (see also Lee et al. 1990, Caputo 1997, and
refs. therein) that the RR Lyrae variables in M92 are highly evolved
HB stars is consistent with the observational results of Storm et
al. (1994) who measured the absolute magnitudes of two variables in
M92 using the Baade-Wesselink technique.  The mean value of
$<\!\!M_{V}(RR)\!\!>$ that they obtained for these two stars is
brighter than the $<\!\!M_{V}(RR)\!\!>$-[Fe/H] relation they derived for
field RR Lyrae variables by the same technique by 0.21$\pm$0.15 mag.
While we predict a smaller offset, our result is within the errors of
the value obtained from this very small sample.

Often observers will not use the few RR Lyrae variables in a blue HB
cluster to set the apparent magnitude of the HB, but will use instead
the reddest non-variable stars on the blue HB.  SHB models (see
fig. 1, 2, \& 3 in Lee et al. 1990) indicate that these stars have
also evolved far from the ZAHB, but somewhat less so than the RR Lyrae
variables.  At best, this practice will remove only partially
the need to take into account HB morphology when estimating the
distance modulus of a cluster.  
Care must also be exercised when turning this problem around and using 
a globular cluster of known distance to set the luminosity of 
RR Lyrae variables.  For example, the globular cluster NGC 6752 has an 
extremely blue HB and lacks RR Lyrae variables.  Its distance 
modulus has been measured from white dwarf fitting (Renzini et al. 1996).
The absolute magnitude of the blue HB stars in NGC 6752 should 
not be assigned without correction to RR Lyrae variables of 
its metallicity. 

Finally, we note that any detailed comparison between theory and
observation is still hampered by empirical uncertainties on cluster
distances as well as on individual cluster metallicities.

\acknowledgments

Partial support for this work was provided by NASA grant NAG5-8406 (P.D.),
NSF grant AST-9319229 (R.Z.), and the Creative Research 
Initiative program of the Korean Ministry of Science \& Technology 
(Y.-W.L. and S.Y.).

\clearpage
 
\begin{table*}
\caption{Parameters used for the HB track construction.} \label{tbl-1}
\begin{center}
\begin{tabular}{ccc}
\tableline
\tableline
$Z$   &  $Y$   &     Core Mass($M_\odot$)\\
\tableline
0.0001 &0.2356 &0.5013\\
0.0004 &0.2370 &0.4981\\
0.0007 &0.2390 &0.4981\\
0.0010 &0.2409 &0.4901\\
0.0040 &0.2416 &0.4877\\
\tableline
\tableline
\end{tabular}
\end{center}
\end{table*}

\begin{table*}
\caption{The slope in the $<\!\!M_{V}(RR)\!\!>$-[Fe/H] relation as a function of metallicity.} \label{tbl-2}
\begin{center}
\begin{tabular}{cc}
\tableline
\tableline
$Z$   &  $\mu$   \\
\tableline
0.0001 -- 0.0004  & 0.154\\
0.0004 -- 0.001  & 0.231\\
0.001 -- 0.004  & 0.234\\
\tableline
\tableline
\end{tabular}
\end{center}
\end{table*}

\newpage

FIGURE CAPTIONS

\figcaption[f1.eps]{
Theoretical calibration, based on the SHB models, of
$M_{V}(RR)$ as a function of HB Type, for each metallicity.\label{fig1}
}
\figcaption[f2.eps]{
Same as Figure 1, but showing the sensitivity of the models
to the choice of $\sigma$.\label{fig2}
}


\figcaption[f3.eps]{
For the case where HB type = 0, the mean absolute visual magnitudes
$<M_{V}(RR)>$ of the RR Lyrae variables is plotted against [Fe/H].
The method of least squares was used to fit the straight line to the
points.  Its equation is:  $<M_{V}(RR)>$ = 0.21[Fe/H]+0.89.\label{fig3}
}

\figcaption[f4.eps]{
For the globular clusters (open circles) in two ranges of
galactocentric distance ($R_{gc}$), [Fe/H] is plotted against HB type.
The dashed line, which is a fit to the data in the upper diagram, is
reproduced in the lower so that the differences between the two groups
of clusters are more apparent.  The x's mark the mean HB types used in
the calculation of the values of $<M_{V}(RR)>$ (see text and Fig. 5).
\label{fig4}
}
\figcaption[f5.eps]{
These diagrams illustrate the differences between the
$<M_{V}(RR)>$-[Fe/H] relations for the globular clusters in the inner
and outer halos.  The points represent our estimates of $<M_{V}(RR)>$
at the values of [Fe/H] and mean HB type (the x's in Fig. 4).  The
solid lines are fits by the method of least squares to the data points
($\mu$ = 0.36 and 0.22 and $\gamma$ = 1.04 and 0.90, for the inner and
outer halo relations, respectively).  The dashed line in each diagram
represents the case where HB type = 0.\label{fig5}
}





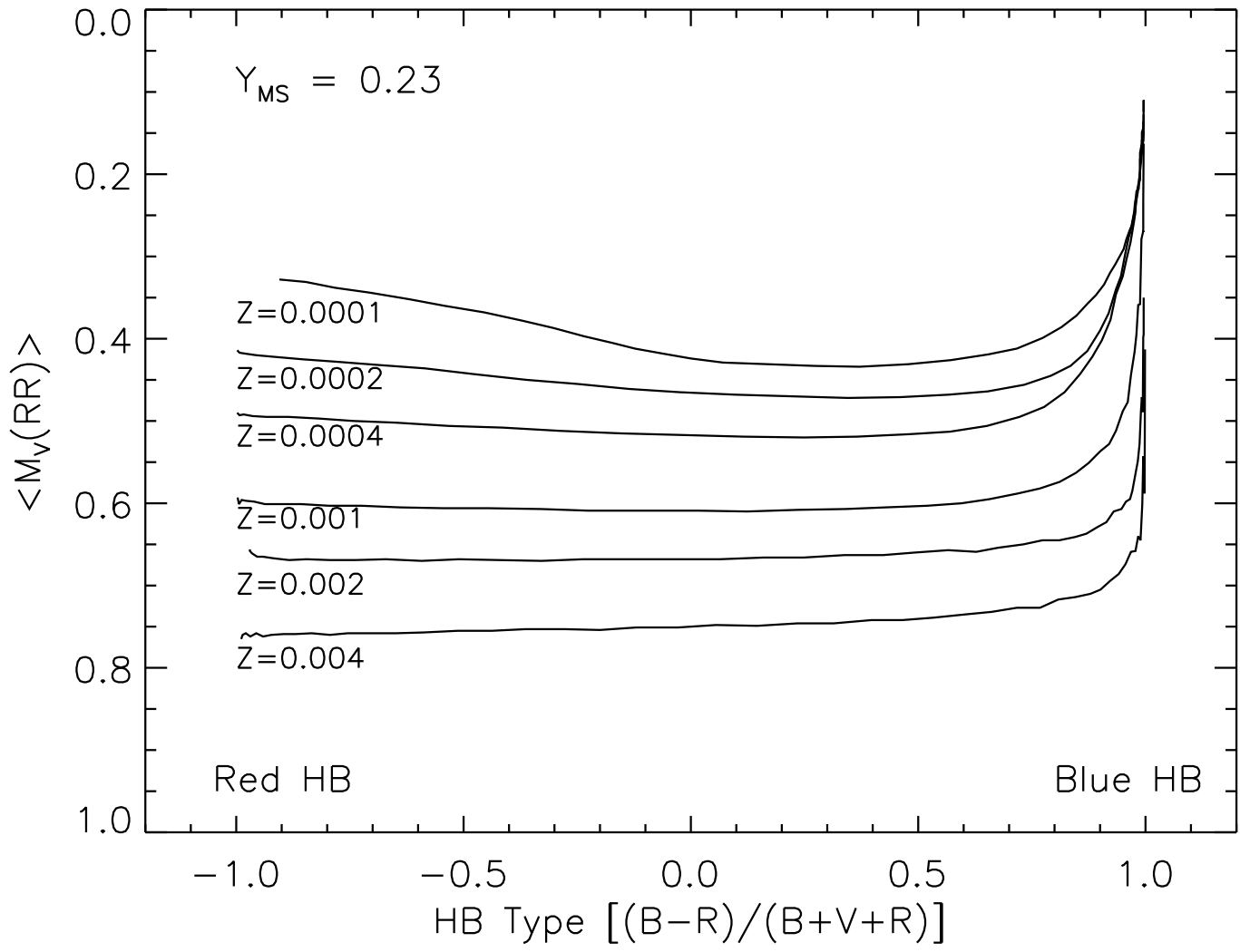
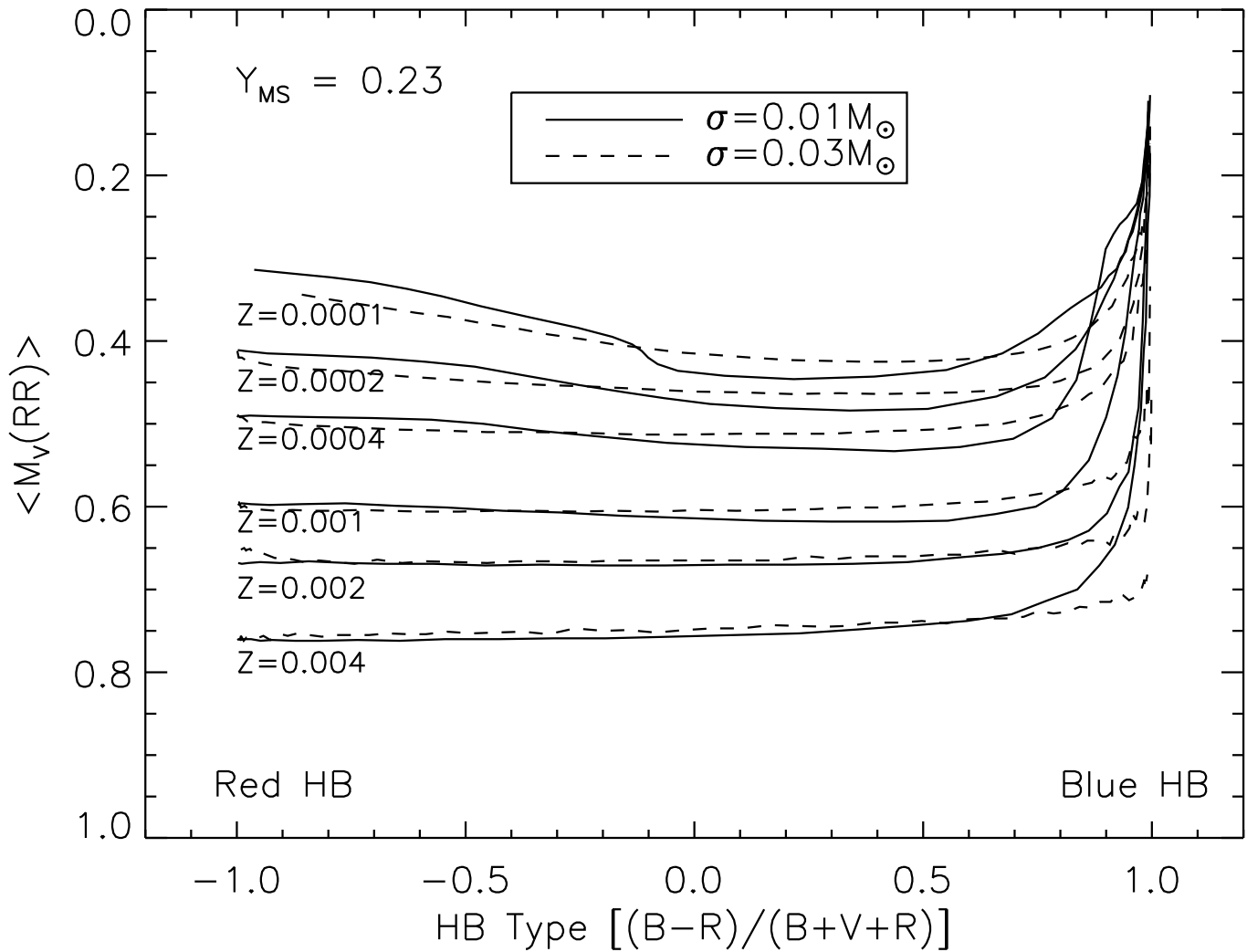
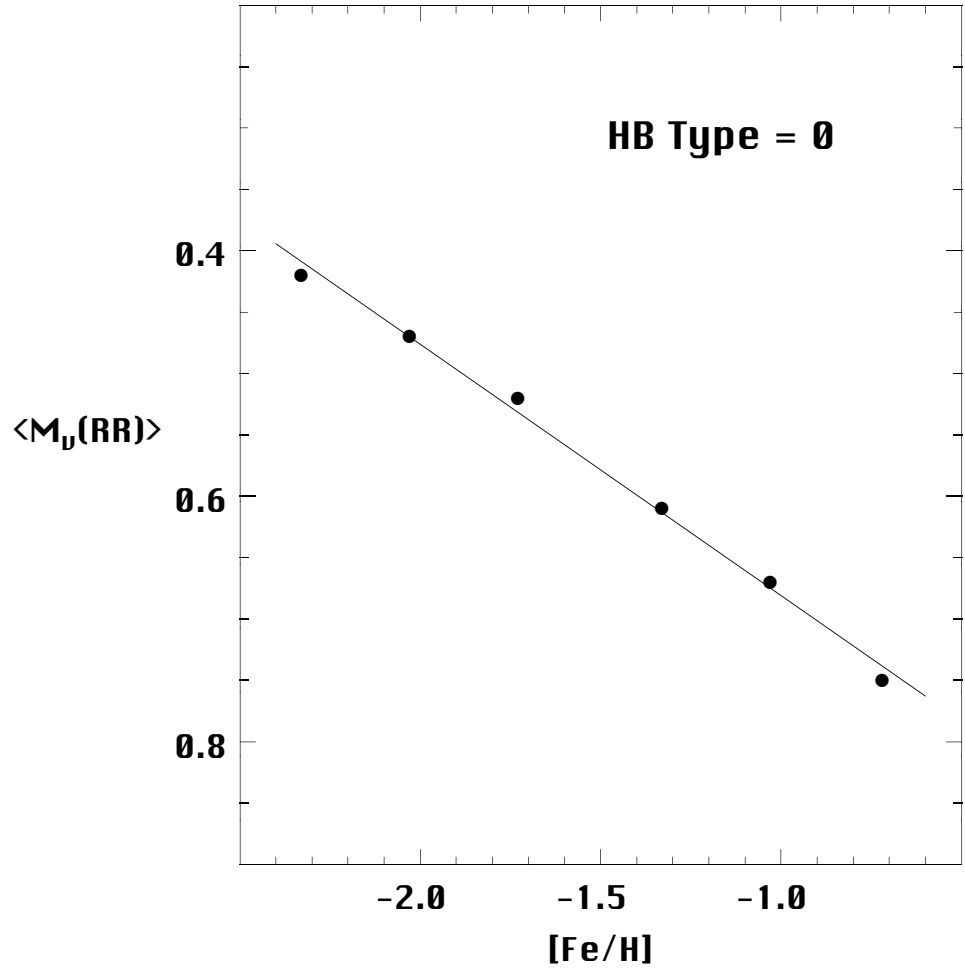
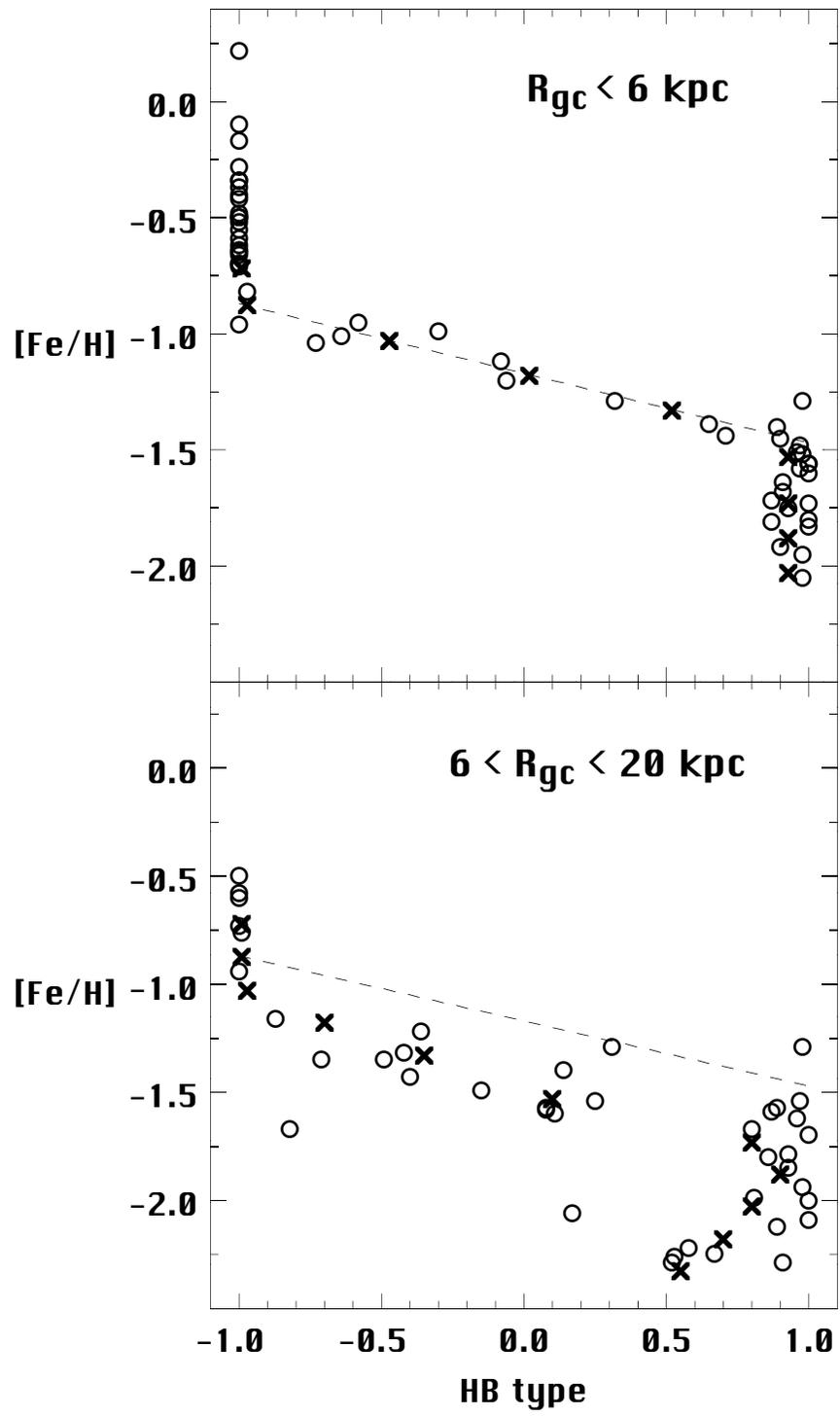
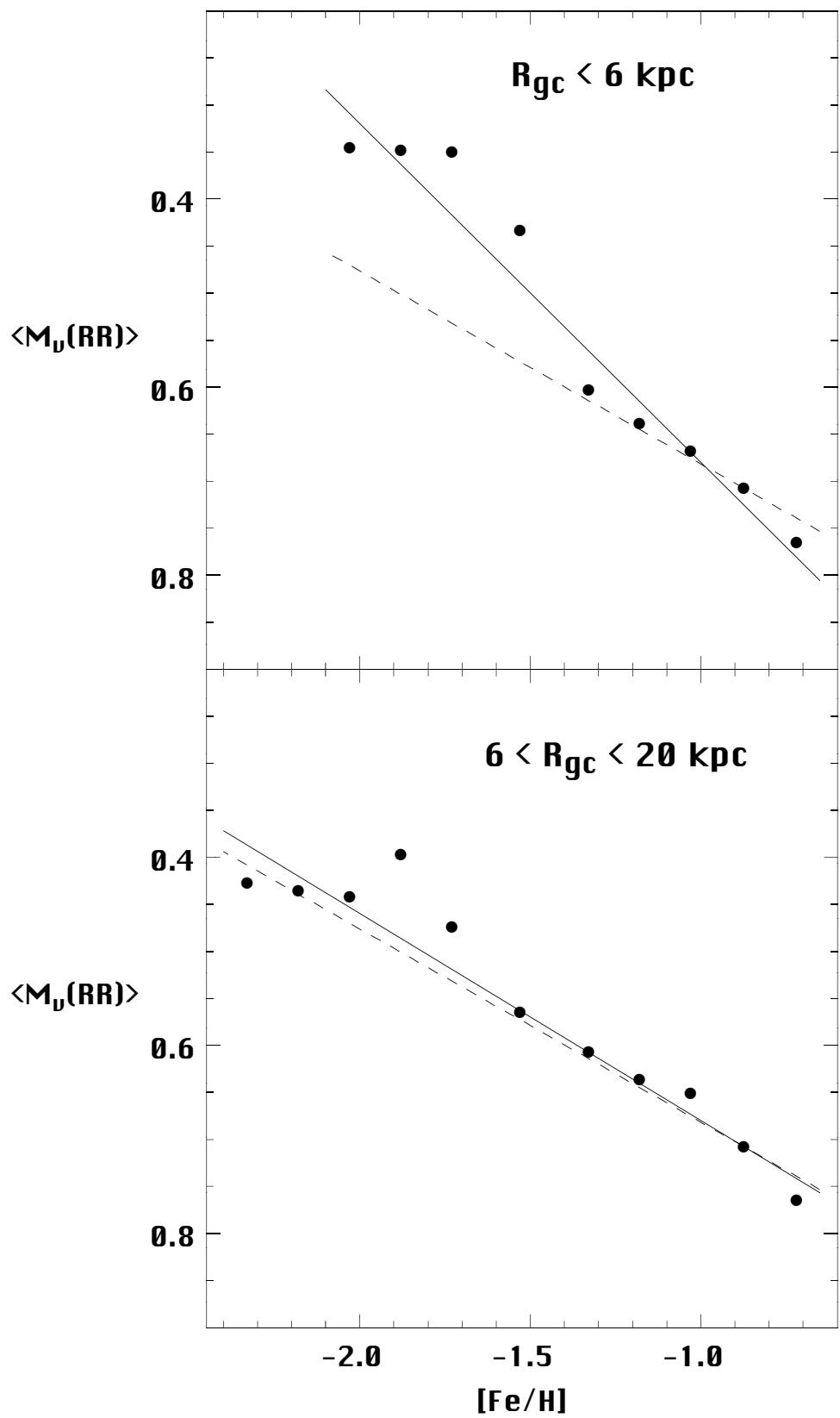
\end{document}